# The Problem of Modeling of Economic Dynamics

S. I. Chernyshov,[1,2] A. V. Voronin,[2*] and S. A. Razumovsky[3**]

[1]*Kharkov National Economic University (Kharkov, Ukraine)*
[2]*"Lemma" Insurance Company (Kharkov, Ukraine)*
[3]*Karkov State Academy of Physical Culture (Kharkov, Ukraine)*
[*]E-mail address: voronin61@ukr.net
[**]E-mail address: rsa_777@mail.ru
(in Russian: http://chvr-article.narod.ru)

**Abstract**

The correctness of Harrod's model in the differential form is studied. The inadequacy of exponential growth of economy is shown; an alternative result is obtained. By example of Phillips' model, an approach to correction of macroeconomic models (in terms of initial prerequisites) is generalized. A methodology based on balance relations for modelling of economic dynamics, including obtaining forecast estimates, is developed. The problems thus considered are reduced to the solution of Volterra and Fredholm integral equations of the second kind.

**Introduction**

Originally, the authors' objective was to analyze the procedure of the construction of differential equations employed in modeling of macroeconomic processes. The results proved to be substantially unexpected, because a number of contradictions were found. In the course of investigations intended to resolve these contradictions, alternative concepts of mathematical modeling of economic processes were formed.

Section 1 is concerned with establishing incorrectness of the widespread in literature Harrod's model, which states an opportunity of macroeconomic growth during unlimited interval of time. Aforesaid incorrectness is caused by applying the instrument of the continuous analysis to the relations whose character is essentially discrete. Investigation of difference Harrod's model brings us to the conclusion, that on the definite range fragment of time a contradiction emerges in the solution.

Correct Harrod's model which argues an inevitability of economic crisis' emerging was formulated in categories of continuous analysis. This model bases on the integral dependence of capital on the intensity of the income and gives constructive opportunities to prevent crisis; first of all by means of a priori estimate of the relevant instant of time. In particularly, it is demonstrated that the activation of the economy presages a crisis. It must be underlined that the discussed crisis and above-mentioned peculiarity of deference solution are agreed in timed.

In section 2 we also demonstrate the incorrectness of Phillips' model: this model has acquired the status of a classical one and is widely represented in special literature, including manuals. A method of correcting this model is proposed, which is based on the above-mentioned approach to the construction of relations between macroeconomic functions of different dimensions: the initial background of economic content is left unchanged at that.

The use of Phillips' "new" model is reduced to the solution of an ordinary differential equation of the second order whose coefficients, in contrast to the "classical" interpretation, are variables. This point substantially widens the spectrum of potentially possible ways of behavior of the economic system, and it seems to be important for practical applications. We give references to literature where versions of analytical solution of the posed problem can be found. A corresponding numerical algorithm is also given.

Arguments that relations of the balance of financial flows are the best fit to the objectives of economic-mathematical modelling form the ideological basis of the content of sections 3 and 4. At the beginning, by an iteration procedure, the static model of value balance is given time dependence, after which the use of a Taylor expansion allows us to derive a system of differential equations. After that, the Cauchy problem is reduced to a system of Volterra integral equations of the second kind: this system has rather favourable, from the point of view of numerical realization, properties.

In this regard, the "balance" is considered as an alternative to unjustified refraction in the economic sphere of the methodology of the construction of mathematical models taken from the field of natural sciences, such as mechanics. As a matter of fact, differential equations of this science organically follow from conditions of equilibrium of an infinitesimal element, whereas analogous constructions in problems of economics have no objective meaning.

Finally, in section 4, we develop an approach to the forecasting of the behavior of an economic system that includes a certain number of participants. The problem is reduced to the solution of the Fredholm integral equation of the second kind whose kernel exclusively depends on factors that characterize relations between the participants. In addition, the free term reflects the index of the cost price of production and planned results of the activity. We propose to refine on the forecast by means of carrying out variational evaluation in the range of changes in the cost price and the results for a resolvent less affected by dynamics.

As tools of such stabilization, there appear mechanisms of an effective interaction between the participants that are objectively inherent in an economic cluster. Here, we employ reasoning related to peculiarities of the solution to the integral equation of the second kind whose kernel depends on the argument, as well as techniques of matrix analysis.

**1. Harrod's model and macroeconomic growth**

Harrod's model of economical development in the representation of L.V. Kantorovich' and A.B. Gorstko [1, pp. 160-161] [#] is defined by the following relations:

$$Y(t) = C(t) + S(t); \ S(t) = I(t); \ S(t) = mY(t);$$
$$d_t K(t) = I(t); \ K(t) = nY(t), \qquad (1.1)$$

where $Y(t)$ is the national income; $C(t)$, $S(t)$ are volumes of consumption and accumulation per year; $I(t)$ is a volume of investment per year; $K(t)$ is the capital.

All mentioned quantities are measured in money equivalent; $d_t = d/dt$, $t$ is the dimensionless time, measuring with year number; $0 < m < 1$ и $n$ is the dimensionless constants (approximately, $m \sim 0.5$; $n \sim 10$). As regards $n$ it is characterized as the number of years during which the yearly income "counterbalances" the capital. The differential equation that formally follows from (1.1), and the solution to this equation take, respectively, the form:

$$d_t K(t) = sK(t), s = m/n;$$

$$K(t) = K_0 e^{st}; I(t) = I_0 e^{st}; Y(t) = Y_0 e^{st}, \qquad (1.2)$$

where $I_0 = sK_0$; $Y_0 = K_0/n$, $K_0 = K(0)$; $I_0 = I(0)$; $Y_0 = Y(0)$.

---
[#] Note that all page references in the text are given according to Russian editions of corresponding literature sources.

As far as volumes of financial flows per year are used in (1.1), it is obvious that these relations are discrete and may be presented in the form:

$$I_n = mY_n; \qquad (1.3)$$

$$d_t K_n = I_n; \qquad (1.4)$$

$$K_n = nY_n, n = 0, 1, 2, ..., \qquad (1.5)$$

where $I_n = I(t_n)$; $Y_n = Y(t_n)$; $K_n = K(t_n)$, $t_n = n$, $n$ is year number.

But the differentiation of the discreetly changing capital $K(t)$ requires the involvement of the instrument of the theory of summarized functions [2] and fraught with difficulties. Really:

$$d_t K(t) = \sum_{j=0}^{n} I_j d(t - t_j), d(t) = \begin{cases} 0, t \neq 0; \\ \infty, t = 0, \end{cases} \int_{-\infty}^{\infty} d(h) dh = 1;$$

therefore, instead of (1.4) it is rational to use the law of the capital forming, which, strictly speaking, is meant:

$$K_n = K_0 + \sum_{j=1}^{n} I_j, \qquad (1.6)$$

where $K_n = K(t_n)$; $I_j = I(t_j)$, $t_j = j$.

From (1.3), (1.5), (1.6) it follows, that

$$K_1 = K_0(1+s); K_2 = K_0(1+s)^2; ...;$$

$$K_n = K_0(1+s)^n = K_0 e^{n\ln(1+s)}; \quad Y_n = (K_0/n)e^{n\ln(1+s)}; \quad I_n = sK_0 e^{n\ln(1+s)}; \qquad (1.7)$$

up to the geometric progression formula

$$\sum_{n=1}^{N} I_n = sK_0 e^{\ln(1+s)} \frac{e^{N\ln(1+s)}-1}{e^{\ln(1+s)}-1}$$

and the comparison of this sum with the capital of $N$ year $K_N = K_n$, $n=N$ demonstrates that the equality between them is reached at value $N$, approximate to $s^{-1}$. In other words, the capital gain during the time from $t=1$ to $t=N$ provided to be equal to the capital $K_N$, which includes initial capital $K_0$ as well.

So we face a contradiction, permitting to make rather substantial conclusions:

– as contrasted to (1.1), the interpretation in difference form (1.3), (1.5) and (1.6), which argue a presence of the peculiarity of solution at $t \sim s^{-1}$ is objectively inherent to the Harrod's model;

– it would be illegal to set in (1.7) $\ln(1+s) \approx s$, because the value $s$ is small, what brings to the solution, which coincides with (1.2);

– such a replacement conceal mentioned peculiarity, radically changing the way of solution (1.7) into stably growing functions $K(t)$, $I(t)$ and $Y(t)$ at the unlimited argument $t$ in (1.2).

Let's notice that income for $n$ years amounts

$$Y_0 + \sum_{j=1}^{n} Y_j = \frac{1}{m}\left(I_0 + \sum_{j=1}^{n} I_j\right), \qquad (1.8)$$

and that because of coefficient $m^{-1} > 1$ this sum grows more fast than (1.6). Obviously, in a certain year $n$, associated with $n$, the sums (1.6) and (1.8) appear equal. Surely in the range of the error of chosen discritization.

So, the solution (1.2) is inadequate to the representation of the macroeconomic dynamics. It caused by the deep contradictoriness of the relation $d_t K(t) = I(t)$ from (1.1). Really, functions $K(t)$ and $I(t)$ are discrete, whereas the derivative $d_t K(t)$ is used in its usual meaning.

From this point of view, arguments of W. Kecs and P. Teodoresky [2, pp. 168-169] concerning problematical character of the use of generalized function in the derivation of differential equations seem to be rather urgent. These functions mostly serve the purpose of simplification of intermediate transformations in the process of solving the problems posed by means of continuous analysis, when the coefficients or the free terms are discontinuous.

So it is needed to constitute the correct analog of Harrod's deference model (1.3), (1.5) and (1.6) in categories of continuous analysis. Obviously, the relation (1.6) may be slightly reduced from the arithmetical law of the capital formation to the integral:

$$K(t) = \int_{-T}^{t} I(h) dh = K_0 + K_R, \tag{1.9}$$

where $T$ is the period of the initial capital $K_0$ accumulation;

$$K_R(t) = \int_{0}^{t} I(h) dh \tag{1.10}$$

is the capital, realized during the period $t > 0$. Accordingly, the derivative

$$d_t K(t) = I(t) \tag{1.11}$$

has its usual meaning.

Function $I(t)$ in (1.9) corresponds to the intensity of the investment flow

$$Y(t) = C(t) + I(t); \tag{1.12}$$

$$I(t) = mY(t) \tag{1.13}$$

(see (1.1), (1.3)), $Y(t)$ and $C(t)$ – also have the meaning of the intensity of flows of the income and consumption measured analogous $I(t)$ in money equivalent, attributed to the unit of dimensionless time $t$. Let's link this unit to the one year, implying that $t$ can running the meanings even of the indefinitely small quantities now.

But the income $Y(t)$ understanding as an intensity of the flow cannot be compared analogously to (1.5) with the capital $K(t)$, which has a money equivalent. Along with it, when $t \to 0$ parameter $n \to \infty$ and the uncertainty emerges.

How can we generalize the relation (1.5) in the case of the continuous time $t$ ? Let's give some considerations about it:

- an income realized in the period of time from 0 to $t$ can be used for the comparison with the capital

$$Y_R(t) = \int_0^t Y(h)dh;$$

- relation (1.5) becomes integral, namely

$$K(t) = n \int_t^{t+1} Y(h)dh, \ t \geq 0;$$

- relation $K(t)/Y_R(t)$, following the logic of "balancing" in (1.5), amounts $n$ when $t=1$, $n/2$ when $t=2$, ..., 1 when $t=n$ ;

- using the same proportion, we get $2n$ when $t = 1/2$, $3n$ when $t = 1/3$, ..., $n/t$ for the arbitrary instant of time $t$ and accordingly when $t \to 0$ a peculiarity emerges and it must be conjugated with $K_0 = nY_0$.

Summarizing the aforesaid, we suppose that

$$K(t) = \frac{n}{t}\int_0^t Y(h)dh; \tag{1.14}$$

and the peculiarity at $t=0$ can be eliminated by the Lopital's rule. The relations (1.11) – (1.14) corresponds a Harrod's model in the continuous interpretation. Putting $Y(t)$ from (1.13) in (1.14) and using (1.11), we get

$$K(t) = \frac{K_0}{1-st}, \tag{1.15}$$

from which

$$I(t) = \frac{I_0}{(1-st)^2}; \ Y(t) = \frac{Y_0}{(1-st)^2}; \tag{1.16}$$

Cauchy problem corresponds to this solution:

$$d_t K(t) - \frac{s}{1-st} K(t) = 0, \ K(0) = K_0;$$

$$d_t I(t) - \frac{2s}{1-st} I(t) = 0, \ I(0) = sK_0; \ Y(t) = \frac{1}{m} I(t) \tag{1.17}$$

and, therefore, considered model practically realizes the above-mentioned peculiarity of the difference solution at $t = N \sim s^{-1}$.

How can we explain the criticality of the relationships (1.15), (1.16) at $t = s^{-1}$? So let us notice, that instead of $t^{-1}$ in (1.14) we can put the function $f(t)$, which satisfy the conditions $d_t f(0) = 1$ and $f(n) = n$:

$$K(t) = \frac{n}{f(t)} \int_0^t Y(h) dh; \tag{1.18}$$

accordingly instead of (1.15)

$$K(t) = \frac{K_0}{1 - sf(t)}$$

and it is obvious that the criticality fit with the solution of the $f(t) = s^{-1}$, can be postponed only by the diminishing of the capital accumulation rate.

The dependence of $m$ on the variable $t$ can be considered in the same way. In other words, an income (1.12) is distributing between the investment and consumption unequally at the different instants of time. In this case the capital is defining by the solution of the Cauchy problem, which follows from the relations (1.11), (1.13) and (1.14):

$$d_t K(t) - \frac{m(t)}{n - tm(t)} K(t) = 0, \ K(0) = K_0. \tag{1.19}$$

But as the consequence of (1.13), when $t = s^{-1}$ the amount of the capital in (1.14) is

$$K(s^{-1}) = \int_0^{s^{-1}} I(h) dh,$$

in other words, there is a coincidence with $K_R(s^{-1})$ from (1.10). Thus, we have

$$K(s^{-1}) = K_R(s^{-1})$$

according to (1.9) in the considered instant of time $K_0 = 0$.

So, at the beginning the total income $Y_R(t)$ reaches the amount equal to the capital $K(n)$ at $t = n$. Because of the further capital growth the total volume of the investment reaches an amount equal to the capital $K(s^{-1})$ at $t = s^{-1}$. At the same time the initial capital $K_0$ transforms into the income $Y_R(t)$ and thus relationships (1.15), (1.16) reverses into uncertainty. Accordingly, the Cauchy problem (1.17) loses its sense at $K_0 = 0$. The situation can be interpreted in terms of the economic chaos, collapse coming, and also in terms of the limitedness of the period of time for forecasting.

In this context let's draw attention to the conditions of the function growth, which follows from (1.15), (1.16):

$$K(n) = K_0/(1-m); \quad I(n) = I_0/(1-m)^2; \quad Y(n) = Y_0/(1-m)^2;$$

so at $m = 0,5$ we get $K(n) = 2K_0$; $I(n) = 4I_0$; $Y(n) = 4Y_0$. Consequently, practically following the approach of the income $Y_0/(1-m^2)$, one can specify the meaning $n$, what is very important for the identification of the crisis moment when $t = n/m$. Really, the parameter $m$ of the model (1.11) – (1.14), in contrast to $n$, is a priori given.

While using more general dependence (1.18) instead of (1.14) the function $f(t)$ fixed on the interval, for example, $t \in [0, n]$ can be extrapolated to the critical instant of time $t_k$, when $f(t_k) = s^{-1}$. Thus, taking the relation (1.14) as a basis (or as a planned growth model), we have an opportunity to define the moment of the crisis coming, guiding also by the specifics of the economic situation development. It is characterized by the function $f(t)$ and by the solution of the problem (1.19).

The noticed circumstances allow to accept, for example, $K(n)$ in the capacity of $K_0$ for the next stage of the economic development. An appropriate process of the organizational rearrangement can be presented in the context of conjugation $n$ with the capital depreciation $K_0$. Let us notice that the depreciation can be easily taking into account by using relationship of the form $(1-at)K(t)$ on the place of function $K(t)$ in (1.9) and (1.14), $a>0$. $K(t)$ is determined in this case by the solution of the Cauchy problem:

$$d_t K(t) - \frac{a+s-2ast}{1-(a+s)t+ast^2} K(t) = 0, \; K(0) = K_0.$$

But the model (1.11) – (1.14) do not take account of the cumulative effect of flowing of the investment into the capital, because in (1.9) they are simply summarized. Let

$$K(t) = \int_{-T}^{t} (1+rh) I(h) dh,$$

where $r>0$ is a constant (one can take $r \sim 0,1$). Then instead of (1.11) one get

$$d_t K(t) = (1+rt) I(t) = m(1+rt) Y(t).$$

Substitution of $Y(t)$ from this relationship to (1.14) leads to the differential equation

$$d_t K(t) - \frac{s(1+rt)}{1-st-srt^2} K(t) = 0,$$

whose solution

$$K(t) = K_0 \left\{ \frac{1}{\sqrt{-srt^2-st+1}} + \left[ \frac{-2st-s-\sqrt{4sr+s^2}}{-2st-s+\sqrt{4sr+s^2}} \right]^{\frac{s}{2\sqrt{4sr+s^2}}} \right\}$$

became uncertain when

$$t = -\frac{1}{2r} + \sqrt{\frac{1}{4r^2} + \frac{1}{sr}}$$

And, obviously, the coming of crisis approaches with increasing $r$.

Returning to the model (1.1) let us notice that at the beginning it was considered in the finite difference statement [3, 4, с. 193-199]. The solution (1.2) emerged as a result of some kind of symbiosis of the continuous and discrete analysis is erroneous. This kind of symbiosis is typical for a great number of the well-known proceedings devoted to the macroeconomic modelling [5 – 7].

## 2. An analysis of macroeconomic models and their correction

Harrod-Domar's model of the development of the economic is presented by P. Allen [6, pp. 75-78] in the following form:

$$Y(t) = C(t) + I(t); \; C(t) = (1-m)Y(t); \; I(t) = n_* d_t Y(t), \quad (2.1)$$

where $Y(t)$, $C(t)$ and $I(t)$ are intensities of the flows of income, consumption and investments, respectively; $0 < m < 1$ is a constant; the constant $n_* > 0$ has the dimension of time; $t$ is dimensional time. The differential equation of the problem and the solution to it has the form

$$d_t Y(t) = (m/n_*) Y(t); \; Y(t) = Y_0 e^{mt/n_*}, \; Y_0 = Y(0), \quad (2.2)$$

respectively.

By virtue of the fundamental dependence

$$d_t K(t) = I(t), \quad (2.3)$$

the last relation in (2.1) is equivalent to the following: $d_t K(t) = n_* d_t Y(t)$, or

$$K(t) = n_* Y(t) + K_0 - n_* Y_0, \; K_0 = K(0). \quad (2.4)$$

Eliminating the functions $Y(t)$ and $I(t)$, we arrive at the equation

$$d_t K(t) - (m/n_*) K(t) = mY_0 - (m/n_*) K_0, \quad (2.5)$$

whose solution has the form

$$K(t) = (K_0 - n_* Y_0)(e^{mt/n_*} - 1),$$

and, consequently, $K_0 = 0$, or $K_0 = n_* Y_0$ in (2.4).

In the first of these two cases, it turns out that the initial capital $K_0$ is absent, whereas the investments $I_0 = I(0)$ and the income $Y_0$ do exist, which contradicts the common sense. In the second case, equation (2.5) becomes homogeneous, and the solution to it is analogous to (2.2):

$$K(t) = n_* Y(t), \qquad (2.6)$$

which also follows from (2.4). However, such dependence contradicts the use of the notion of the infinitesimal. Indeed, the intensity of income for $t = t_i$ is defined as follows:

$$Y(t_i) = \frac{1}{t_*} \int_{t_i - 0{,}5t_*}^{t_i + 0{,}5t_*} Y(h) dh,$$

where $t_*$ is a small interval of time. Accordingly, by (2.6), the capital is

$$K(t_i) = n \int_{t_i - 0{,}5t_*}^{t_i + t_*} Y(h) dh, \; n = \frac{n_*}{t_*}. \qquad (2.7)$$

In other words, $K(t_i)$ represents the value of income averaged over the interval $t_*$, and, as is obvious, the dimensionless parameter $n \to \infty$ for $t_* \to 0$.

Simultaneously, the difference between the notions of income am the intensity of income disappears at $t = t_i$. Thus, by (2.7), relation (2.6) can be considered only for a finite period $t = t_i$, and, as a consequence, it has essentially discrete character. Moreover, the use of the dimensionless time $t = t/t_*$, alongside with $n$, transforms relations (2.1) – (2.3) and (2.6) into the model [1, pp. 160-161] that, as shown in section 1, is incorrect.

An analogous situation occurs in Phillips' simplest model [6, p. 79]:

$$Z(t) = (1 - m) Y(t); \; r d_t Y(t) = Z(t) - Y(t),$$

where $Z(t)$ is the flow of demand for the product; $r$ (per time unit) is the constant of a lag between demand and production. As a matter of fact, here again two functions of different dimensionality are conjugated by means of a coefficient: namely, an intensity of the flow and the rate of its change.

Phillips' general model is defined [6, pp. 81-82] by the equations

$$d_t Y(t) = l \left[ I(t) - m Y(t) \right]; \; d_t I(t) = k \left[ n d_t Y(t) - I(t) \right], \qquad (2.8)$$

where the constants and their units of measurement are the following: $k > 0$, 1/unit of time; $n > 0$, unit of time; $0 < m < 1$; $l > 0$, 1/unit of time. The problem is reduced to solving an ordinary differential equation of the second order with constant coefficients:

$$d_t^2 Y(t) + a d_t Y(t) + b Y(t) = 0, \ a = k + ml - nkl; \ b = mkl, \ t \geq 0. \tag{2.9}$$

By virtue of (2.3), under the condition

$$d_t K(0) = k[nY(0) - K(0)],$$

which is an analog of the relation that follows from the solution (2.5), equation (2.8) takes the form

$$d_t Y(t) = l[d_t K(t) - mY(t)]; \ d_t K(t) = k[nY(t) - K(t)]. \tag{2.10}$$

(Note that exactly in this manner the considered model is treated by A. Bergstrom [7, pp. 40-41].)

The arguments concerning (2.6) are directly extended to the second of these equations. Accordingly, the solution to equation (2.9) does not represent an adequate reflection of the dynamics of macroeconomic development. An analogous situation also occurs for other models [6, 7]. In each case, in this or that way, there exists conjugation of the capital with an intensity of the income via a dimensional coefficient.

Following the methodology of section 1, we represent equation (2.10), relating the capital to the flows $d_t K(t)$ and $Y(t)$, in the following form:

$$K(t) = \frac{1}{t} \int_e^t L(h) dh, \ L(t) = nY(t) - \frac{1}{k} d_t K(t), \ 0 < e < t, \tag{2.11}$$

where $e$ is a small quantity. In other words, we have used a correct approach to the formation of the capital at the expense of corresponding flows of intensities by means of integration. Here, $t^{-1}$ plays the role of a proportionality factor that relates $K(t)$ to the capital accumulated by means of $Y(t)$ and $d_t K(t)$ during the period of time from $0$ to $t$.

Thus, adhering to the idea [6], put into relations of the type (2.6) and (2.10), we practically realize it on an arbitrary interval of time, including $t = e \to 0$, when the L'Hôpital rule comes into play.

From (2.11), it follows that

$$K(t)=\frac{nk}{1+kt}\int_{e}^{t}Y(h)\,dh;\ d_tK(t)=-\frac{nk^2}{(1+kt)^2}\int_{e}^{t}Y(h)\,dh+\frac{nk}{1+kt}Y(t).$$

As a result of simple transformations, including a passage to the limit $e \to 0$, the first of relations (2.10) takes the form of the differential equation (2.9), but now its coefficients are variables:

$$a(t)=ml+\frac{2k-nkl}{1+kt};\ b(t)=\frac{2mkl}{1+kt}.$$

This equation can be represented as follows:

$$d_t^2 Y(t)+\left(a+\frac{b}{t}\right)d_t Y(t)+\frac{g}{t}Y(t)=0,\ t\geq 1,\qquad(2.12)$$

where $t=1+kt$ is a dimensionless variable of time $t=1+kt$; the coefficients $a=ml/k$, $a=ml/k$ and $g=2ml/k$ are dimensionless. It can be reduced to the solution of the degenerate hypergeometric equation [8, p. 392, №2.120; pp. 428-431]. The well-known substitution [9, с. 130]

$$Y(t)=u(t)\exp\left[0.5(a-at-b\ln t)\right]$$

transforms equation (2.12) into the following:

$$d_t^2 u(t)+c(t)u(t)=0,\ c(t)=-\frac{a^2}{4}+\frac{2g-ab}{2t}+\frac{b(2-b)}{4t^2},\ t\geq 1,\qquad(2.13)$$

from which, for

$$p=-a^2/4;\ r=(2g-ab)/2;\ s=b(2-b)/4,$$

we get

$$t^2 d_t^2 u(t)+\left(pt^2+rt+s\right)u(t)=0,\ t\geq 1.$$

The solution to this equation can be obtained in a closed form [8, p. 392, №2.154; pp. 547-548]. Note also the well-known substitution [9, p. 131] that reduces (2.13) to the Riccati equation $d_t u(t)+u^2(t)+c(t)=0$; however, for a given function $c(t)$, there are no constructive methods of its solution.

To evaluate the function $Y(t)$ that satisfies (2.12) under the conditions $Y(1)=Y_1$ and $d_t Y(1)=Y_1'$, one can capitalize on the known procedure of the reduction of such a problem to a Volterra integral equation of the second kind (see, e.g., [10, pp. 16-18]). Indeed, from the notation $d_t^2 Y(t) = j(t)$ it follows that

$$Y(t) = \int_1^t (t-h) j(h) dh - (1-t) Y_1' + Y_1,$$

and after substitution into (2.12) we get the equation

$$j(t) = \int_1^t k(t,h) j(h) dh + q(t), \, t \geq 1, \tag{2.14}$$

where the kernel and the free term are defined by the following expressions:

$$k(t,h) = -a - b - \frac{b-h}{t}; \, q(t) = -a - \frac{b-(1-t)}{t} Y_1' + \frac{g}{t} Y_1.$$

The solution of equation (2.14) can be found with the help of the procedure of successive approximations, under a practically arbitrary choice of the initial element $j_0(t)$ [11]:

$$j_{n+1}(t) = \int_1^t k(t,h) j_n(h) dh + q(t), \, t \geq 1; \, n = 0, 1, \ldots. \tag{2.15}$$

**3. Economic-mathematical model based on price balance**

Phillips' model of macroeconomic dynamics contains four sufficiently abstract parameters: the rate of reaction $k$ (an inverse of a constant lag of investments); an investment coefficient $n$ (an index of the accelerator's power); a multiplier $m$ that characterizes a part of the income directed to investment; the rate of the influence of the production output on the demand [6].

The evaluation of the intensity of the income is reduced here to solving the differential equation (2.9) that is in wide use in engineering. For example, it describes free oscillations of a mass suspended on a spring, under the condition of viscous resistance. All the parameters of such a system, including external forces, are extremely concrete and can be measured. The differential equation is derived strictly on the bases of fundamentals of mechanics [12, c. 43-49].

In this regard, one should bear in mind two points. The first one is that the spectrum of possible solutions to the above-mentioned equation is objectively insufficient for an adequate representation of macroeconomic functions. As a result, the attention of economists was attracted by equations of nonlinear theory: see, in particular, the arguments of T. Puu [13, p. 7]. However, their interpretations in categories of an objective sphere are rather problematic.

At the same time, we have shown above that Phillips' model in the interpretation [6] is incorrect. Using ideological prerequisites of this model, we have reduced the problem to the solution of a differential equation whose coefficients are time-dependent. Owing to this fact, the class of solutions has become much wider, but uncertainty in the choice of the above-mentioned parameters has remained.

The second point concerns the fact that economics does not contain laws for idealized objects that could be put in correspondence to a material point. However, economics, in its turn, has advantage over mechanics, which is embodied in the equation of the balance of financial flows. From this point of view, possibilities of mathematical modeling in economics and mechanics can be characterized as having different orientations.

Let us turn to a static system of balance equations $x = Ax + c$, or

$$x_i = \sum_{j=1}^{n} a_{ij} x_j + c_i, \ i = 1, 2, ..., n, \tag{3.1}$$

where $x_i$ is the cost of the product of the $i$-th participant (in the case of invariable production volumes it is analogous to the price); $a_{ii}$ is the part of the cost of the product of the $i$-th participant that constitutes his income; $a_{ij}$ is the part of the cost of the product of the $j$-th participant consumed by the $i$-th participant; $c_i$ personal contribution of the $i$-th participant (including remuneration of labor, payment for materials and outside services, etc.); $t \geq 0$ is dimensional time. The coefficients $a_{ij}$ and the free terms $c_i$ are assumed to be given.

As in such a situation, except for different extraordinary factors (see below) $a_{ij}$, $c_i \geq 0$, and, obviously, the sums of the elements of each line of the matrix $A$ do not exceed unity, whereas at least one of these sums is less than unity, we have: $\|A\|<1$ [14, pp. 329-331]. Accordingly, the quantities $x_i$ can be determined by means of successive approximations:

$$x_{s+1} = Ax_s + c, \ s = 0, 1, ... \tag{3.2}$$

[15, pp. 120-121].

A point of principle is that the system of equations (3.1) can be attached dynamic character by setting $x = x(t)$, and

$$x_s(t) = x(t_s); \ x(t_{s+1}) = x(t_s + t_*), \tag{3.3}$$

where $t_*$ is a sufficiently small interval of time. In this regard, just the first step of the process (3.2) – (3.3) from each point $t_s = t$ of the considered interval would suffice to achieve the set goal.

Thus, there appears the relation

$$x_i(t+t_*) = \sum_{j=1}^{n} a_{ij} x_j(t) + c_i, \ t \in [0, t_*], \tag{3.4}$$

and retaining in the Taylor expansion of $x_i(t+t_*)$, say, three terms, after a transition to a dimensionless time variable $t = t/t_*$, we arrive at the following differential equation:

$$d_t^2 x_i(t) + 2 d_t x_i(t) + 2 x_i(t) = 2 \sum_{j=1}^{n} a_{ij} x_j(t) + 2 c_i, \ t \in [0, 1]. \tag{3.5}$$

By use of the initial conditions $x_i(0) = p_i$ and $d_t x_i(0) = p_i'$ (the constants $p_i$ and $p_i'$ are assumed to be given) the problem is reduced to the solution of a system of Volterra integral equations of the second kind with respect to the functions

$$j_i(t) = d_t^2 x_i(t), \tag{3.6}$$

from which it follows:

$$x_i(t) = \int_0^t (t-h) j_i(h) dh + p_i' t + p_i, \ i = 1, 2, ..., n. \tag{3.7}$$

Indeed, upon the substitution of expressions (3.6) and (3.7) into (3.5), we get

$$j_i(t) = l \sum_{j=1}^{n} \int_0^t k_{ij}(t,h) j_j(h) dh + q_i(t), \; i = 1, 2, ..., n, \tag{3.8}$$

with the parameter $l = 2$; the kernels are given by

$$k_{ij}(t,h) = \begin{cases} a_{ij}(t-h), \; j \neq i; \\ (a_{ij}-1)(t-h)-1, \; j = i; \end{cases}$$

the free terms are

$$q_i = \begin{cases} 2\left[\sum_{j=1}^{n} a_{ij}(p'_j t + p_j) + c_i\right], \; j \neq i; \\ 2\left[\sum_{j=1}^{n} a_{ij}(p'_j t + p_j) - p'_j(1+t) - p_j + c_i\right], \; j = i. \end{cases}$$

The solution to the system of equations (3.8) is obtained by means of successive approximations, i.e.,

$$j_{i,s+1}(t) = l \sum_{j=1}^{n} \int_0^t k_{ij}(t,h) j_{j,s}(h) dh + q_i(t); \; j_{i,0} = 0, \; s = 0, 1, ...; \tag{3.9}$$

$$x_{i,s}(t) = \int_0^t (t-h) j_{i,s}(h) dh + p_{di} t + p_i, \; i = 1, 2, ..., n, \tag{3.10}$$

or it can be represented as a series expansion in powers of $l$ whose terms contain sums of integrals with iterated kernels $k_{ij}(t,h)$ [11, pp. 59-61].

However, suppose that the above-mentioned requirements to $a_{ij}$ and $c_i$ are not fulfilled. For example, they can take on negative values which reflects payment of debts, subventions, use of stocks as well as other factors of this kind. If the sum of the elements of the $j$-th column of the matrix $A$ exceeds unity, it means that the $j$-th participant sells the product to his partners at a price higher than the real value. In any case, we will consider the assumption $\|A\| < 1$ to be invalid.

In this situation, we introduce in (3.1) the notation $I - A = B$, where $I$ is a unity matrix. Now, the solution to the equation $Bx = c$ is obtained with the help of the following process of successive approximations [16, pp. 70-73]:

$$x_{s+1} = (I - rB'B)x_s + aB'c, \ s = 0, 1, \ldots$$

where $B'$ is the transpose of the matrix $B$; $0 \leq a \leq 2/\|B'B\|$.

All the arguments and transformations related to (3.2) – (3.10) still hold. Only the values of $a_{ij}$ and $c_i$ change as, accordingly, the kernels and the free terms of equations (3.9) do. It should be noted that this fact does not influence practically the procedure of numerical realization. The same concerns retaining in (3.5) derivatives of higher order: with increasing the length of the interval $t_*$, any a priori estimates in this sense are difficult to make.

It is not difficult to take into account in (3.1) a lag between sales of the product and production of the form $x_j(t) = x_j(t + b_j)$. The enumerated possibilities characterize a substantial advantage of the proposed approach over the solution of the considered problem in the differential form (3.5).

Another advantage of integral equations is due to a possibility to extend transformations to the case when the coefficients $a_{ij}$ and the free terms $c_i$ depend on time, which rather important in the context of further consideration. Indeed, algorithms of integration of piecewise continuous bounded functions are rather universal (see, e.g. [17]), and, from this point of view, the presence of $a_{ij} = a_{ij}(h)$ and $c_i = c_i(t)$ in (3.8) would not be of fundamental importance.

## 4. Forecasting of the development of an economic situation

Naturally, the participants of an economic system are interested in prospects of further activity. In this regard, we assume that they forecast in some way the dynamics of their mutual relations and of external demand (directly related to the cost of production) as well as the price level by the end of a considered period of time. Therefore, the functions $a_{ij}(t)$, $c_i(t)$, and the constants $x_i(1) = r_i$ are known, and we are faced with solving a boundary-value problem for the system of equations (3.5) under conditions on $x_i(t)$ for $t = 0$; $t = 1$. Using, by analogy with the previous case, (3.6), we get:

$$x_i(t) = \int_0^t (t-h) j_i(h) dh - t\int_0^1 (1-h) j_i(h) dh + (r_i - p_i)t + p_i. \qquad (4.1)$$

Upon the substitution of this expression into (3.5) the problem is reduced to the solution of a system of Fredholm integral equations of the second kind:

$$j_i(t) = l \sum_{j=1}^n \int_0^1 k_{ij}(t,h) j_j(h) dh + q_i(t), \; t \in [0, 1]; \; i = 1, 2, \ldots, n, \qquad (4.2)$$

with the parameter $l = 2$; the kernels are defined by the equations

$$k_{ij}(t,h) = \begin{cases} t(h-1)a_{ij}(h), \; t \leq h \leq 1; \\ (t-1)h a_{ij}(h), \; 0 \leq h \leq t, \; j \neq i; \end{cases}$$

$$k_{ij}(t,h) = \begin{cases} [t+1-t a_{ij}(h)](1-h), \; t < h \leq 1; \\ (t-2)h a_{ij}(h), \; 0 \leq h \leq t, \; j = i \end{cases}$$

(the second of these equations is discontinuous at the diagonal $h = t$); the free terms are

$$q_i(t) = \begin{cases} 2\left\{\sum_{j=1}^n a_{ij}\left[(r_j - p_j)t + p_j\right] + c_i\right\}, \; j \neq i; \\ 2\left\{\sum_{j=1}^n (a_{ij} - 1)\left[(r_j - p_j)t + p_j\right] - r_j + p_j + c_i\right\}, \; j = i. \end{cases}$$

For the purpose of finding the functions $j_i(t)$ that satisfy equations (4.2), a number of methods of numerical realization were developed [16]. At the same time, the system of equations (4.2) can preliminarily be reduced to a single Fredholm integral equation of the second kind [10, pp. 77-78]:

$$\Phi(t) = l \int_0^n K(t,h) \Phi(h) dh + Q(t), \; t \in [0, n], \qquad (4.3)$$

where the sought function, the free term and the kernel have the form

$$\Phi(t) = j_i(t - i + 1); \; Q(t) = q_i(t - i + 1); \; K(t,h) = k_{ij}(t - i + 1, h - j + 1),$$

$i = 1, 2, \ldots, n$.

A solution to this equation does exist and is unique for $l \neq l_h$, where $l_h$, $h = 1, 2, \ldots$ are characteristic numbers of the kernel $K(t,h)$. After the evaluation of the functions $j_i(t)$, the solution to the problem is obtained by means of substitution of these functions into (4.1).

For $l = l_h$, the homogeneous equation

$$\Phi(t) = l\int_0^n K(t,h)\Phi(h)dh; \ h = 1, 2, ..., t \in [0, n] \tag{4.4}$$

has nontrivial solutions. As is obvious, an economic system should avoid such kind of critical regimes of functioning, because they are not favorable for its participants. The means for this is an increase of the efficiency of relations of mutual partnership that is implicitly related to optimization of the coefficients $a_{ij}(t)$.

Certainly, both the cost function $c_i(t)$ and the results of the activity of the participants $r_i$ can be known only approximately. Therefore, to estimate the behavior of an economic system, it is reasonable to carry out variational calculations. In other words, equation (4.3) will be solved repeatedly for a given kernel and under variations of the free term. For this reason, the following representation of the solution seems to be rather useful [11]:

$$\Phi(t) = Q(t) + l\int_0^n R(t,h,l)Q(h)dh, \ t \in [0, n], \tag{4.5}$$

where $R(t,h,l)$ is the resolvent of the kernel $K(t,h)$. To construct the resolvent, one can use the constructive algorithm of S. G. Mikhlin [18, pp. 210-221].

From a formal point of view, the coefficients $a_{ij}(t)$ should also be varied; however, in such a case a forecast of the system's behavior can become practically unrealizable because of a large number of variants.

From this point of view, a role of such an organizing economic system as a cluster is of great importance. Indeed, its participants build up their interrelations on principles of supplying each other with reliable information and coordinate their activity on the basis of criteria of a systematic level. These facts substantially facilitate more objective determination of the function $a_{ij}(t)$.

Retaining (for better transparency of the economic situation) during the forecast period the stability of the coefficients $a_{ij}(t)$, a cluster can then redistribute incomes of the participants inside its organization. A scheme of such redistribution is preliminarily coordinated at an informal level.

Note that the outlined approach to forecasting organically matches the essence of cluster methodology: see the seminal works by M. Porter and a number of other sources [19, 20].

Let us turn to a meaningful side of the parameter $l$ in equation (4.3). At a glance, it is used only for the sake of convenience in the treatment of symbols. This point of view is justified to a certain extent, but, at the same time, one should not the origin of $l=2$. Initially, this coefficient appeared in the procedures of the construction of the system of equations (3.8) and (4.2), and it is due to the Taylor expansion in (3.4):

$$x(t+t_*) = x(t) + \frac{1}{1!}d_t x(t) + \frac{1}{2!}d_t^2 x(t) + \frac{1}{3!}d_t^3 x(t) + \dots .$$

As we retained the first three terms, we got $l=2$. In the case of four terms, one would have $l=6$ and so on. However, with an increase of the interval of the Taylor series, the forecast interval $t_*$ to which the functions $a_{ij}(t)$, $c_i(t)$ and the constant $r_i$ are attached also objectively increases. Consequently, there exists an internal relation between them and the parameter $l$, which, however, cannot be expressed in functional terms. One can just state that, in reality, in (4.2) and, further, in (4.3), (4.4) we have:

$$a_{ij}(t) = a_{ij}(t, l); \ c_i(t) = c_i(t, l); \ r_i = r_i(l).$$

Accordingly, instead of (4.3), we get the equation

$$\Phi(t) = \int_0^n K(t, h, l) \Phi(t) dt + Q(t), t \in [0, n], \qquad (4.6)$$

whose kernel depends on the parameter. As is pointed out by V. I. Smirnov, in the consideration of such equations, one can encounter substantial deviations from Fredholm's theory.

Tamarkin's theorem states that, for certain kernels $K(t, h, l)$ that depend analytically on $l$, the resolvent $R(t, h, l)$ in (4.5) does not exist for any values of this parameter [21, pp. 130-132]. In other words, equation (4.6) proves to be unsolvable (see also [10, p. 49]). We emphasize that the above-mentioned arguments are of qualitative character because of the absence of functional dependence of $K(t, h, l)$.

At the same time, the solvability of equation (4.6) and that of the initial system of equations (3.1) are, obviously, mutually related. Thus, the values $l = l_h$ in (4.4) depend in some way on characteristic numbers of the matrix $A$, whereas the insolvability of equation (4.6) is caused by the closeness to zero of its determinant $\det(I - A)$.

As regards this issue, here emerges a constructive verification of the solvability of equation (4.6), which is based on an investigation into the matrix $A$ with $a_{ij} = a_{ij}(t)$. Namely, the functions $a_{ij}(t)$ should be chosen in such a way that $\det[I - A(t)]$ should not vanish, and the matrix $I - A(t)$ should not be ill-defined (see, e.g., [17]) for all $t \in [0, t_*]$.

Note that ill-definedness implies, in this case, an inadequate overreaction to small perturbations both of functions fulfilled by the links and of the indices of their structural conjugation. In general, for effective functioning of such a system, including reliability of the forecast, it is desirable that the matrix $A(t)$ should satisfy the conditions of Perron-Frobenius' second theorem [22, pp. 247-248]: i.e., it should be nonnegative and indecomposable. In other words, all $a_{ij}(t) \geq 0$, and the directed graph corresponding to the matrix $A(t)$ should be strongly connected.

The latter condition means that any two vertices of the directed graph should possess a directed path that connects them [23, pp. 129-130]. However, in this case, the number of contacts in each pair of the participants proves to be rather large, which, generally speaking, is not typical of the processes of material and financial flows.

However, an advantage of the cluster exactly consists in the fact that an active exchange of experience and knowledge takes place between its participants. By definition, the cluster is characterized by a high degree of ramification of intellectual flows that, although realized on non-repayable basis, can, nevertheless, be represented in money equivalent, which thus ensures indecomposability of the matrix $A(t)$, $t \in [0, t_*]$.